\newif\ifSC
\SCtrue
\ifSC
\documentclass[onecolumn,draftclsnofoot,12pt]{IEEEtran}
\else
\documentclass[conference]{IEEEtran}
\fi
\IEEEoverridecommandlockouts
\usepackage{cite}
\usepackage{amsmath,amssymb,amsfonts}
\usepackage{algorithmic}
\usepackage{xcolor}
\usepackage{amssymb}
\usepackage{bbm}
\usepackage{amsmath}
\usepackage{bm}
\usepackage{amssymb,mathtools}
\usepackage{amsmath}
\usepackage{amsbsy}
\usepackage[font=small]{subcaption}
\usepackage{graphicx} 
\usepackage{pdfpages}
\usepackage{varioref}
\usepackage{xcolor}
\usepackage[font=small]{subcaption}
\usepackage[font=small]{caption}
\usepackage[update,prepend]{epstopdf}
\usepackage{graphicx} 
\usepackage{bbm}
\usepackage{amssymb}
\usepackage{amsmath,amsthm}
\usepackage{color}
\usepackage{amsfonts}
\usepackage{graphicx}
\usepackage{verbatim}
\usepackage{epstopdf}
\usepackage{cite}
\usepackage{tabulary}
\usepackage{mathtools}
\usepackage{enumitem}

\newcommand{\set}[1]{\mathsf{#1}}

\newcommand{\palmexpect}[1]{\mathbb{E}^0{\left[#1\right]}}

\newcommand{\palmexpectx}[2]{\mathbb{E}^{#2}{\left[#1\right]}}

\newcommand{\dd}{\mathrm{d}}

\newcommand\expect[1]{\mathbb{E}\left[#1\right]}
\newcommand\prob[1]{\mathbb{P}\left[#1\right]}

\newcommand{\Ball}{\mathcal{B}}

\newcommand{\ie}{{\em i.e.}~}

\newcommand{\SThres}{\tau}

\DeclareMathOperator*{\expectop}{\mathbb{E}}

\newcommand{\areaballintersect}[4]{\mathcal{A}\left(#2,#3,#4\right)}

\newcommand{\UUCFB}[1]{\overline{\overline{\mathcal{M}_{#1}}}(r_\set{K})}

\newcommand{\LLCFB}[1]{\underline{\underline{\mathcal{M}_{#1}}}(r_\set{K})}

\newcommand{\UCFB}[1]{{\overline{\mathcal{M}_{#1}}}(r_\set{K})}

\newcommand{\LCFB}[1]{{\underline{\mathcal{M}_{#1}}}(r_\set{K})}

\newcommand{\reqEnergy}[1]{\mathcal{E}_{#1}}

\newcommand{\y}{{\mathbf{y}}}
\newcommand{\x}{{\mathbf{x}}}
\newcommand{\expU}[1]{e^{#1}}

\newcommand{\expS}[1]{\exp{\left(#1\right)}}

\def\home{\hbox{\kern3pt \vbox to13pt{}%
   \pdfliteral{q 0 0 m 0 5 l 5 10 l 10 5 l 10 0 l 7 0 l 7 5 l 3 5 l 3 0 l f
               1 j 1 J -2 5 m 5 12 l 12 5 l S Q }%
   \kern 13pt}}

\newcommand{\BoolP}{\mathrm{P}}

\newcommand{\BoolM}{\mathrm{M}}

\newcommand{\BoolT}{\mathrm{T}}
\newcommand{\CFB}[1]{\mathcal{M}_{#1}(r_\set{K})}

\newcommand{\totallambda}[1]{\lambda_{#1}}
\newcommand{\parentlambda}{\lambda_\pt}

\def\BibTeX{{\rm B\kern-.05em{\sc i\kern-.025em b}\kern-.08em
		T\kern-.1667em\lower.7ex\hbox{E}\kern-.125emX}}
\newcounter{relctr} 
\everydisplay\expandafter{\the\everydisplay\setcounter{relctr}{0}} 

\newcommand\numeq[1]%
{\stackrel{\scriptscriptstyle(\mkern-1.5mu#1\mkern-1.5mu)}{=}}
\AtBeginDocument{}

\renewcommand{\x}{\mathbf{x}}
\renewcommand{\y}{\mathbf{y}}
\newcommand{\z}{\mathbf{z}}

\newcommand{\tf}{\mathbf{t}}

\newcommand{\St}{\mathsf{S}}
\newcommand{\1}{\mathbbm{1}}
\newcommand{\pt}{\mathrm{p}}
\newcommand{\drm}{\mathrm{d}}

\newcommand{\ob}{\mathrm{o}}

\newcommand{\dv}{\mathrm{d}}
\newcommand{\bt}{\mathbf{b}}
\renewcommand{\bt}{\Ball}
\newcommand{\B}{\mathcal{B}}
\newcommand{\K}{\mathsf{K}}

\newcommand{\kb}{\lambda_{\mathrm{d}}}

\newcommand{\Abf}[3]{\mathbf{A}(#1)}

\renewcommand{\gg}[2]{\frac{\alpha(r)}{2 \lambda_\drm}}
\newcommand{\C}{\mathcal{C}}

\DeclareMathOperator\erf{erf}

\newcommand{\matern}{Mat\'ern~}

\newcommand{\ggsq}[2]{\frac{\alpha^2{r}}{4 \lambda_\drm^2}}

{}
{}

\newcommand{\commst}[1]{{\color{red}#1}}
\newcommand{\commadd}[1]{{\color{green}#1}}

\renewcommand{\commst}[1]{}
\renewcommand{\commadd}[1]{}
\newtheorem{theorem}{Theorem}

\newtheorem{corollary}{Corollary}
\newtheorem{remark}{Remark}
\newcommand{\insertnotationtable}{	
	\begin{table}
			\caption {\small Notation Table} \label{tab:title}
		\begin{center}
			\begin{tabular}{|p{.5in} | p{2.43in} |}
				\hline 
				\textbf{Symbol} & \textbf{Definition}  \\ \hline 
				$\bt(\x,r)$ & Ball of radius $r$ centred at location $\x$.\\ \hline
				$\Phi$ & 2D process such as MCP or TCP modeling location of sensors. \\ \hline
				$\totallambda{\BoolP}$ & Intensity of PPP.\\ \hline
				$\lambda_\pt$ & Intensity of parent point process.\\ \hline					
				$\lambda_\drm$ & Intensity of daughter point process in a cluster process.\\ \hline
				$r_\K$ & The size of an event $\K$, which may grow with time.\\ \hline
				$R$& The fixed sensing radius associated with each sensor.\\ \hline				
				$\mathsf{S}_j^{(i)}$ & Compact disk of radius $R$ of $j$-th point associted with $i$-th parent and models the sensing zone of sensor.\\\hline
				$\Psi$ & Occupied region in $\mathbb{R}^2$ by all the points. Represents the area falling under the sensing zone of all sensors. \\ \hline
				$\oplus$ & Minkowski addition.\\ \hline
				$y=\|\y\|$ & $L$-2 norm of $\y$.\\ \hline	
				$\Phi_{\drm}^{(i)}$ & Daughter point process coressponding to $i$-th parent.\\ \hline	
				$ \areaballintersect{}{x}{r_\drm}{r}$& Intersecting area between $\bt(\ob,r_\drm)$ and $\bt(\x,R)$: $|\bt(\ob,r_\drm)\cap\bt(\x,R)|$.\\ \hline						
			\end{tabular}
		\end{center}
		
\end{table}}

\allowdisplaybreaks 
\begin{document}
\title{On the Coverage Performance of  Boolean-Poisson Cluster Models for Wireless Sensor Networks}

\author{\IEEEauthorblockN{ Kaushlendra Pandey, Abhishek Gupta}
	\thanks{The authors are with 
		the Modern Wireless Networks Group at
		 the Indian Institute of Technology Kanpur, Kanpur (India) 208016 (Email: kpandey@iitk.ac.in, gkrabhi@iitk.ac.in).}
}

\maketitle
\begin{abstract}
In this paper, we consider wireless sensor networks (WSNs) with sensor nodes exhibiting clustering in their deployment. We  model the coverage region of such WSNs by Boolean Poisson cluster models (BPCM) where sensors nodes' location is according to a Poisson cluster process (PCP) and each sensor has an independent sensing range around it. We consider two variants of PCP, in particular \matern and Thomas cluster process to form Boolean \matern and Thomas cluster  models. We first derive the capacity functional of these models. Using the derived expressions, we  compute the sensing probability of an event  and compare it with sensing probability of a WSN modeled by a Boolean Poisson model where sensors are deployed according to a Poisson point process. We also derive the  power required for each cluster to collect data from all of its sensors for the three considered WSNs. We show that a BPCM WSN has less power requirement in comparison to the Boolean Poisson WSN, but it suffers from lower coverage, leading to a trade-off between per-cluster power requirement and the sensing performance. A  cluster process with desired clustering may provide better coverage while maintaining low power requirements.
\end{abstract}
\IEEEpeerreviewmaketitle
\section{Introduction}
In WSNs, sensors are deployed over a region such as forest or wetlands, to form a wireless network and exchange mutual data to sense an event.  WSN may have a central hub to facilitate the joint detection. There are two essential aspects of WSNs. The first aspect is the coverage aspect \ie to maximize the region covered by sensors, termed the coverage or sensing region. This will ensure that at least one sensor can detect the target event with a certain probability. The second aspect is to minimize energy consumption as wireless sensors have a limited power budget.  Sensors can form small clusters with each cluster having one head, which acts as a gateway to the central hub \cite{iyengar2016distributed}. In this hierarchical network, sensors transmit their sensing data to their local cluster heads, which then communicates it to the central hub to jointly make sensing decision. Such clustering can reduce the power requirement of nodes, but can degrade overall coverage. The deployment of sensors in a WSN is generally random. Hence, the tools of stochastic geometry can be applied to model and analyze WSNs. 
One popular process to model the coverage area of a WSN is the Boolean-Poisson process. The Boolean-Poisson process is defined as the union of independent random objects with their centers located according to a Poisson point process (PPP) \cite{haenggibook,liu2004study,baek2007spatial}.  The random objects denote the individual coverage region of sensors while the centers denote sensors' locations. Owing to the mathematical tractability of PPPs, Boolean-Poisson process is simple yet powerful to derive performance metrics of WSNs such as the probability that a location is not covered, and the expected area of uncovered  region \cite{BaccelliBook,chiu2013stochastic,flint2017wireless}.
The capacity function of the Boolean-Poisson process, which characterizes the sensing probability of an event, was studied in \cite{pandey2018modeling}. 
As the underlying process of the Boolean Poisson model is PPP, the location of sensors nodes is independent of each other in this model. In some scenarios, the deployment of sensors is not entirely independent and the sensors may exhibit clustering in their deployment. This may be due to the easiness in deploying sensors in small groups or to facilitate the communication between the sensor and its gateway by decreasing their mutual distance. 
The Poisson cluster process (PCP) can be used to model the locations of sensors in such scenarios \cite{mekikis2018connectivity}. The two important variants of PCP are the \matern cluster process (MCP) and Thomas cluster process (TCP).  The characterization of contact and nearest neighbor distance distribution for these processes is presented in \cite{pandey2019contact,afshang2017nearesttcp}. To model the coverage region of WSNs exhibiting such clustering, we propose to use  Boolean Poisson cluster models (BPCM) where the underlying process to model sensors' locations is a PCP and each sensor has an independent sensing region around it. There has been limited work to characterize BPCM {\em e.g.} \cite{last1999empty}. However, coverage and sensing performance of WSNs that are deployed according to BPCM has not been studied in detail.\\
In this paper, we consider three WSNs that are deployed according to MCP, TCP, and PPP, respectively. The coverage area of these WSNs can be modeled using Boolean MC, Boolean TC, and Boolean Poisson models (or processes). We first derive the capacity functional of Boolean MC and TC models. Using these expressions, we then compute the sensing probability of an event with a compact spread area. We also provide simple bounds for Boolean MC model to help derive insights for the system. We also derive the power required for each cluster to collect data from all of its sensors for the three considered WSNs. 
Finally, we perform a comparative analysis of these three deployments. We show that clustering decreases the coverage area and sensing probability, especially in the case of sensors with large individual sensing regions. However, it also reduces the power requirement of sensors. In scenarios where sensors have limited power, clustered deployments can provide better coverage performance.
\section{System Model}
In this paper, we consider a wireless sensor network deployed over $\mathbb{R}^2$ space. The locations of the sensors are modeled by a point process $\Phi$ with density $\lambda$. Each sensor has a sensing range around it denoted by $\St_i$ and assumed to be independent of other sensors. The total covered region (\ie the region which falls inside the sensing region of at least one sensor) is given as
\begin{align*}
\Psi= \bigcup_{\z_i \in \Phi} \z_i+\St_i,
\end{align*}
which is known as a Boolean Process/Model and is a special case of Germ-grain model. Each point is termed as a germ with its sensing region as its grain.

\subsection{Sensor network}
 We assume that sensors follow clusterization where the network is made from many cluster heads with each cluster head responsible to control and communicate with sensors assigned to it. Such network can be modeled using a cluster process.
A cluster process consists of daughter point process centered at their parents whose locations are also according to a point process. Let $\Phi_{\pt}=\{\x_i: \forall i \in \mathbb{N}\}$, be a parent point process where $\x_i$ is the location of $i$-th parent point (models the location of cluster center or cluster head in WSN) in $\mathbb{R}^2$. For each point $\x_i$, there is an associated daughter point process $\Phi^{(i)}_{\drm}=\{\y_j^{(i)}:\forall j\in \mathbb{N}\}$, where $\y_j^{(i)}$ is the location of $j$-th daughter point. 
The absolute location of these points are given as $\z_{ij}=\x_i+\y_j^{(i)}.$ Each daughter point process is independent and identically distributed. Now the PP modeling the sensors' location is the union of all  these daughter points \ie
\begin{align*}
\Phi&=\bigcup_{\x_i\in \Phi_{\pt}}\{\x_i+\Phi_{\drm}^{(i)}\},\\
&=\left\{\z_{ij}:\z_{ij}=\x_i+\y_j^{(i)},\x_i\in\Phi_{\mathrm{p}},\y_j^{(i)}\in\Phi_{\drm}^{(i)} \,\forall i,j \right\},
\end{align*}
and known as the cluster process. It is clear from the above discussion that the cluster head is the parent point to all sensors of the cluster. It is intuitive to keep the cluster head at the center of the cluster and as close as possible to the sensors in the cluster to minimize the energy required in communication. We now consider three PPs to model the locations of WSN:
\subsubsection{\matern cluster process}
In MCP, $\Phi_{\pt}$ is a homogeneous PPP  with intensity $\lambda_\pt$. Each  $\Phi_{\drm}^{(i)}$ is  a finite PPP within a ball $\B(\ob,r_\drm)$. The mean number of points in each daughter point process is $m$ and therefore the intensity $\lambda_\drm(\y)$, of each daughter point process will be $\frac{m}{\pi r_\drm^2}\1(||\y||\leq r_\drm)$. Total density of the PP is $\totallambda{\mathrm{M}}=\parentlambda m$.
\subsubsection{Thomas cluster process}
In  TCP, $\Phi_{\pt}$ is a homogeneous PPP with intensity $\lambda_\pt$.  Each  $\Phi_{\drm}^{(i)}$ is  a non-uniform PPP with the intensity
\begin{align*}
\lambda_{\drm}(\y)=\frac{m}{2\pi\sigma^2}\exp\left(-\frac{y^2}{2\sigma^2}\right),
\end{align*}
where  $m$ is the mean number of points in each daughter point process. Total density of the PP is $\totallambda{\mathrm{T}}=\parentlambda m$.

\insertnotationtable
\subsubsection{Poisson point process}
In this, sensors are located according to a PPP with intensity $\totallambda{\BoolP}=\lambda_\pt m$. For the sake of consistency, another independent PPP with intensity $\lambda_\pt$ defines the location of cluster heads. Sensors form clusters by selecting the closest cluster head.

\begin{figure*}[ht!]
	\centering
	\begin{subfigure}[t]{.3\linewidth}
		\includegraphics[width=1\textwidth]{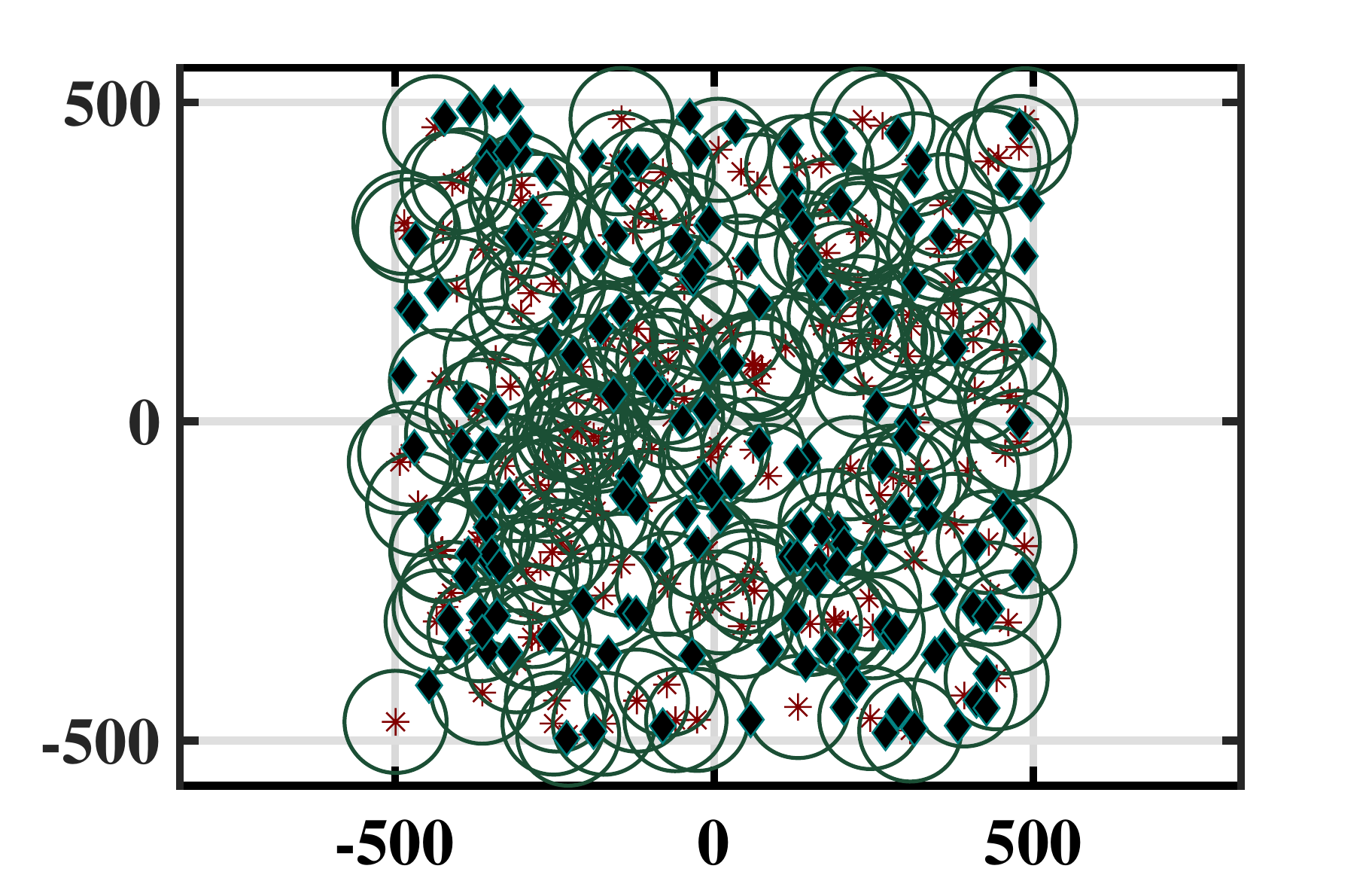}
		\caption{PPP, $R=80$. FAC=$67.9\%$} \label{PPP}
	\end{subfigure}
	\begin{subfigure}[t]{.3\linewidth}
		\includegraphics[width=1\textwidth]{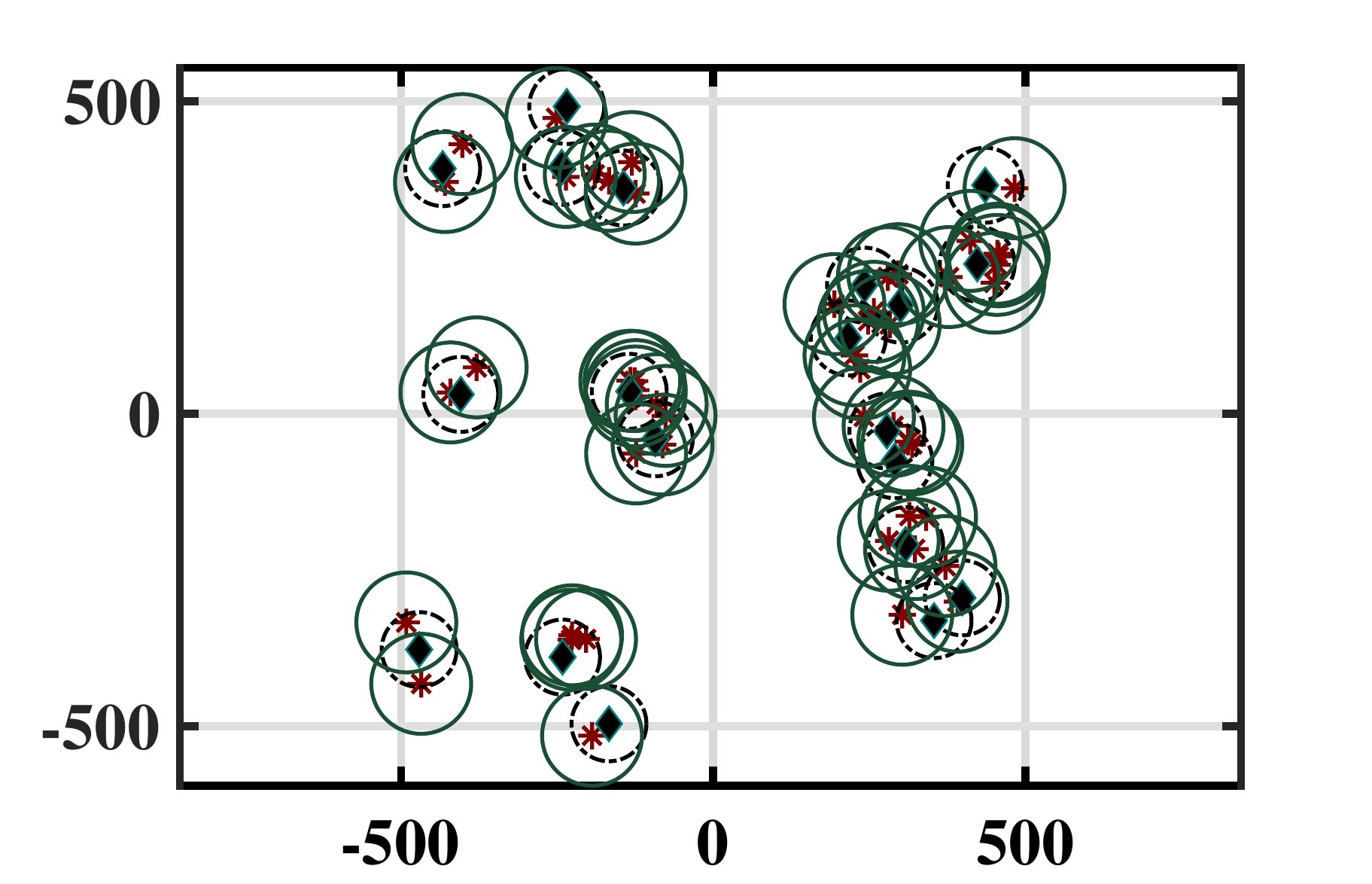}
		\caption{MCP, $R=80$. FAC=$41.4\%$}\label{MCPPP}
	\end{subfigure}
	\begin{subfigure}[t]{.3\linewidth}
		\includegraphics[width=1\textwidth]{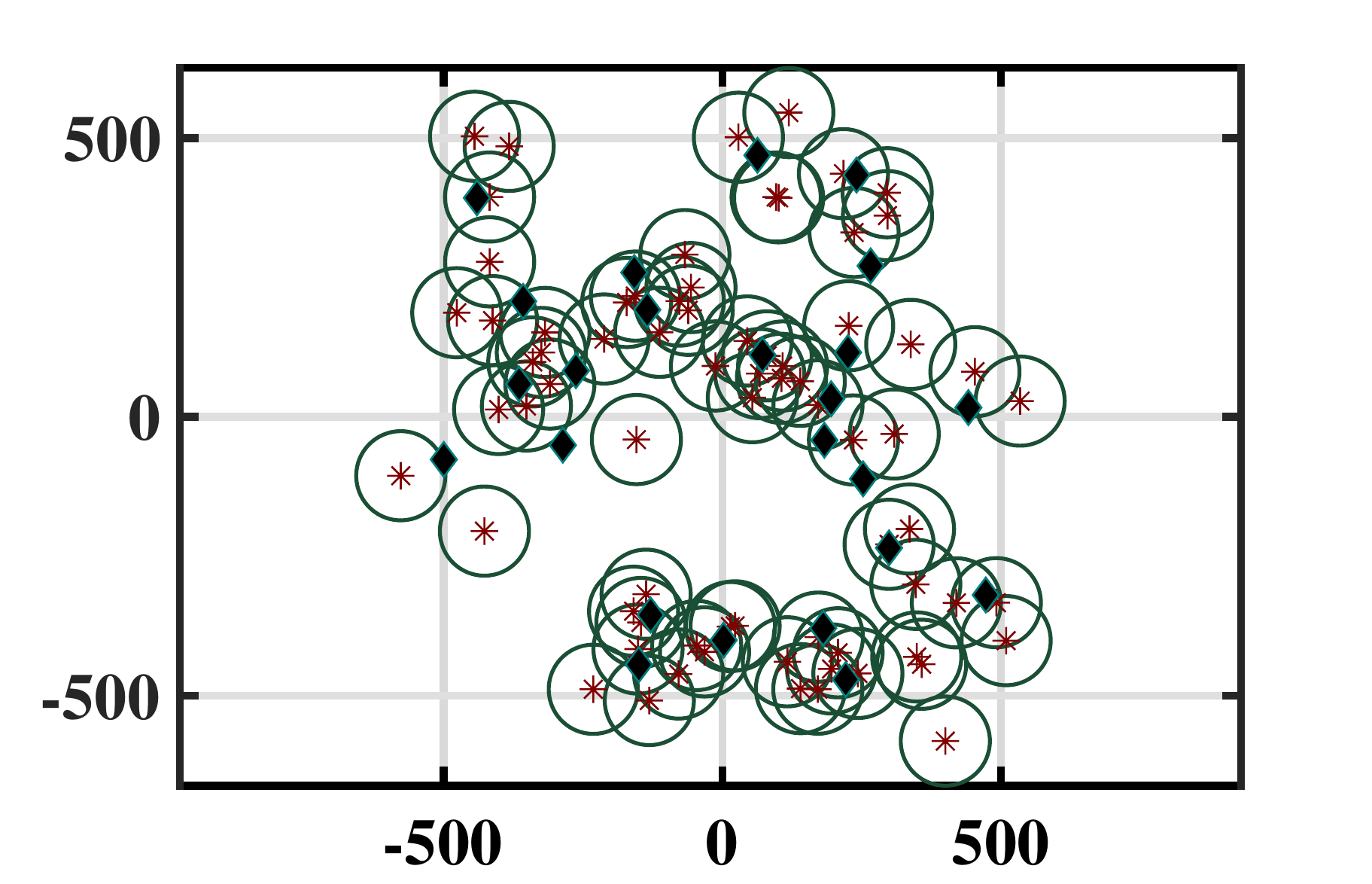}
		\caption{TCP, $R=80$. FAC=$50.7\%$}\label{TCPPP}
	\end{subfigure}  
	\begin{subfigure}[t]{.3\linewidth}
		\includegraphics[width=1\textwidth]{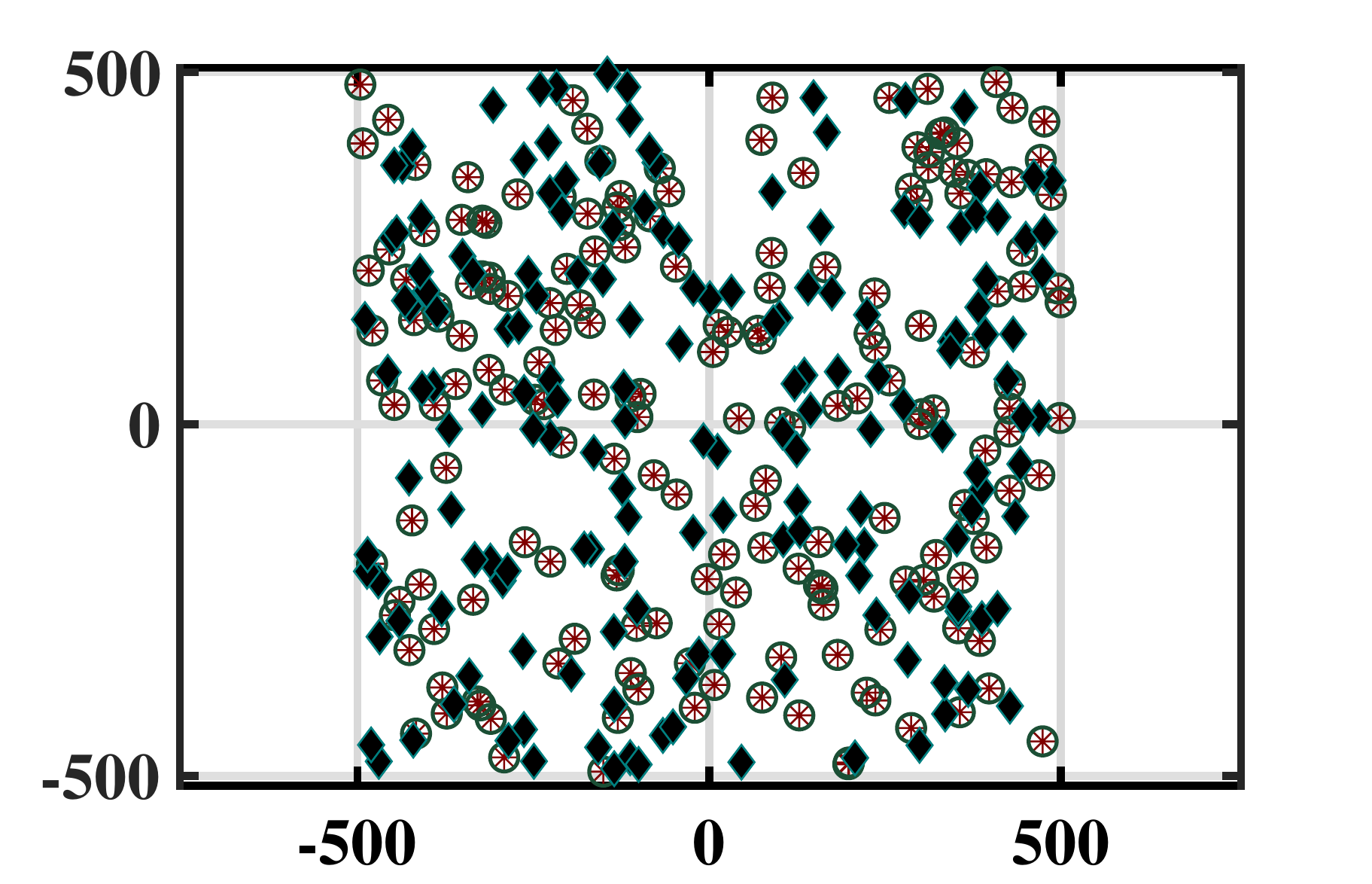}
		\caption{PPP, $R=20$. FAC=$7.8\%$} \label{pppv2}
	\end{subfigure}
	\begin{subfigure}[t]{.3\linewidth}
		\includegraphics[width=1\textwidth]{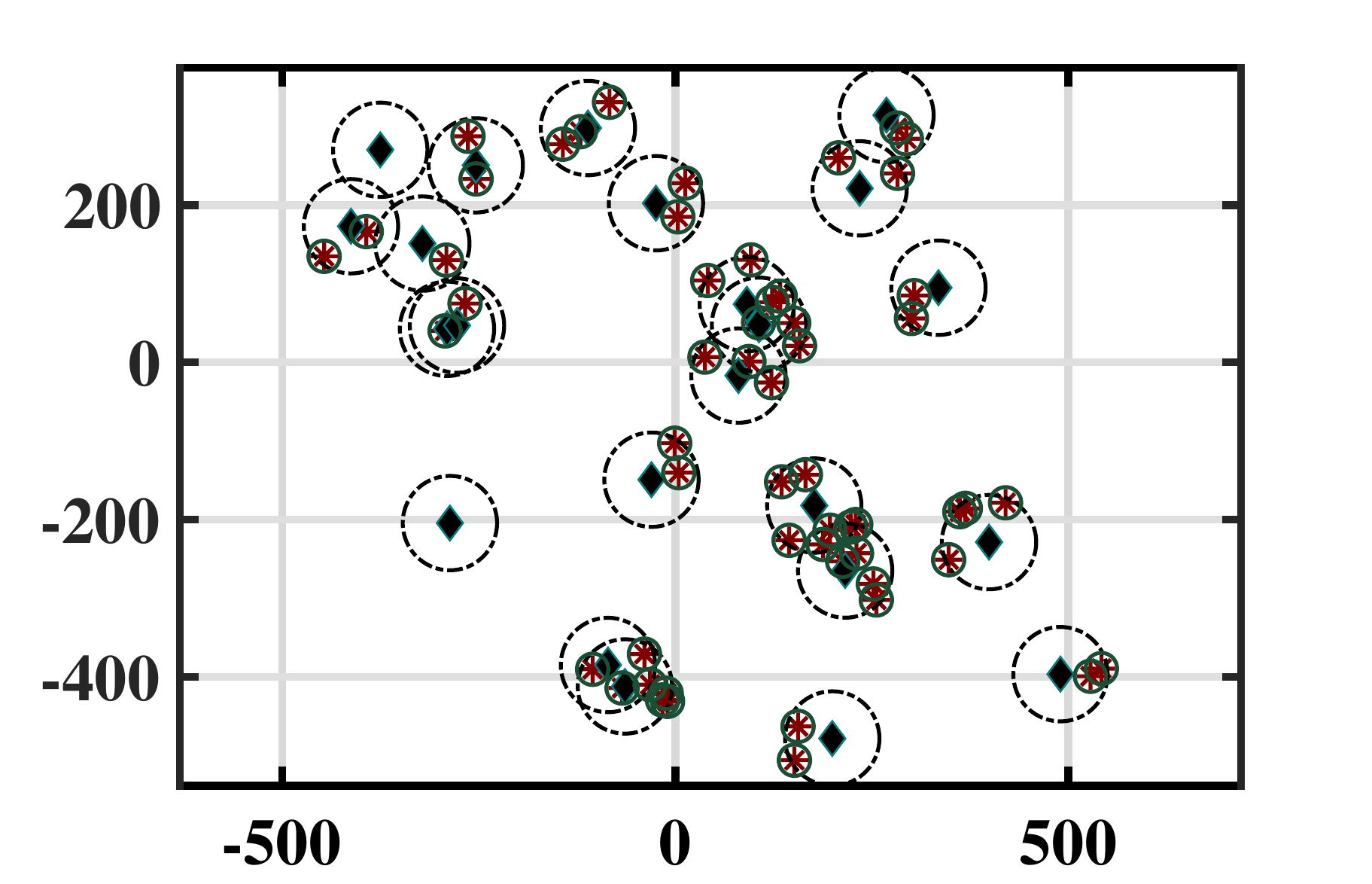}
		\caption{MCP, $R=20$. FAC=$5.4\%$}\label{mcpv2}
	\end{subfigure}
	\begin{subfigure}[t]{.3\linewidth}
		\includegraphics[width=1\textwidth]{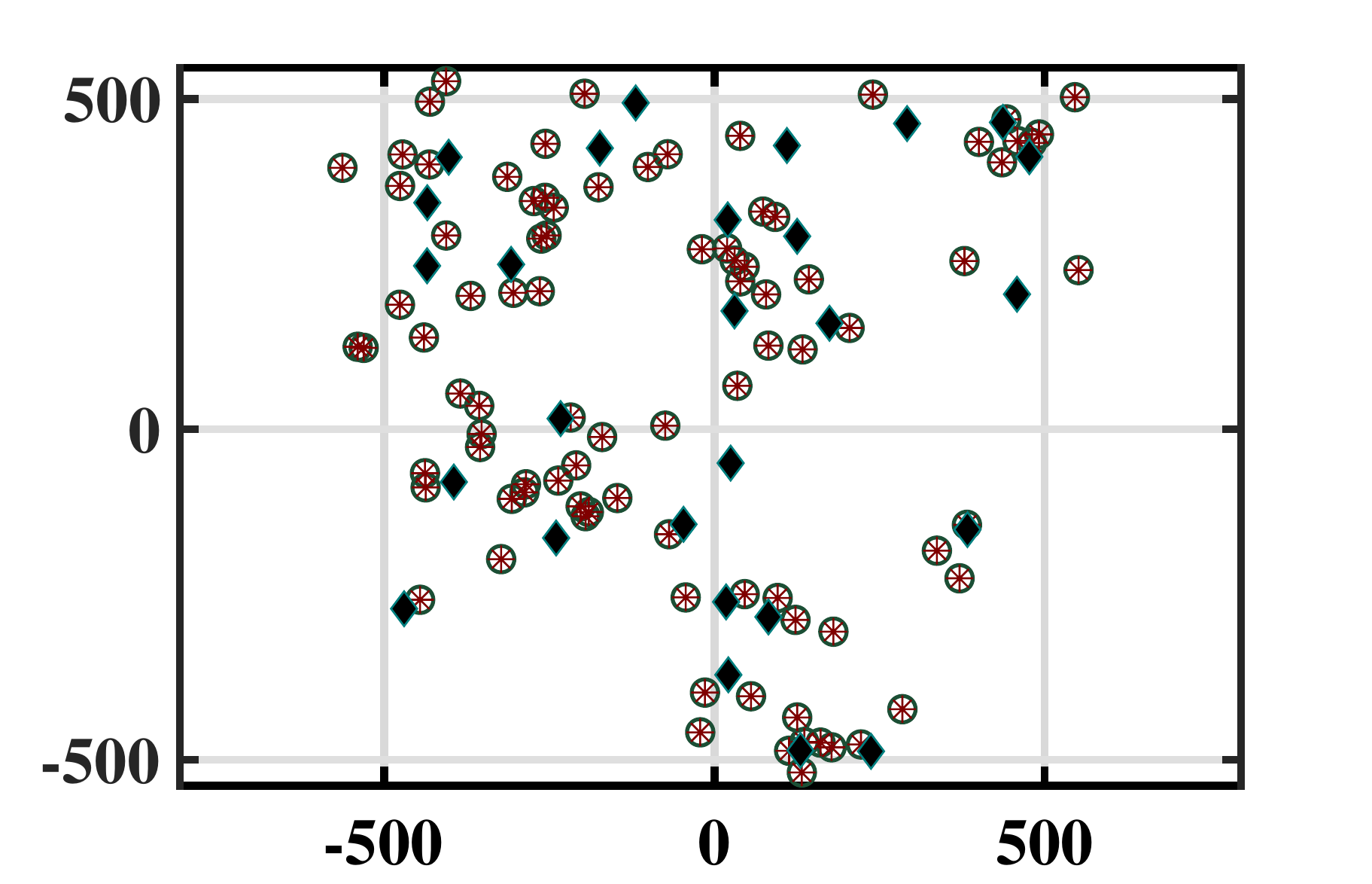}
		\caption{TCP, $R=20$. FAC=$6.8\%$}\label{tcpv2}
	\end{subfigure} 
	\caption{Comparison of the fractional area covered (FAC) by three processes. For the clustered processes $m=3$, $\lambda_{\pt}=20\times10^{-6}$. In Fig. \ref{PPP} the intensity $\lambda_{\mathrm{P}}$ of $\Psi_\BoolP$ is $m\lambda_\pt$. The Fig. \ref{MCPPP} and \ref{mcpv2} shows the $\Psi_\BoolM$ with $r_\drm=60$. The centers of black circle are the parent point of $\Psi_\BoolM$. The points inside the black circle are the respective daughter points. In Fig. \ref{TCPPP} the $\sigma=60$. The region falls under the green circles comes under the sensing region. It is clear from the simulation that for higher values of $R$, $\Psi_\BoolP$ have the highest and $\Psi_\BoolM$ have the lowest coverage. In case of smaller value of $R$, $\Psi_\BoolT$ and $\Psi_\BoolP$ provide the similar coverage area.} \label{fig:fig4}
\end{figure*}
Now, depending on the deployment of sensors, we consider the following processes to model the total coverage region which is the union of all sensing regions. In each model, sensors have their  individual sensing region  $\St_i$ as $\B(\ob,R)$ around it independent of other sensors. Here $R$ is the fixed sensing range of each sensor.
\begin{enumerate}
	\item \textbf{Boolean MC process:} The sensors' locations follow MCP.
	\item \textbf{Boolean TC process:} The sensors' locations follow as TCP.
	\item \textbf{Boolean P process:} The sensors' location are modeled as PPP. Results for this case are known, however this case is considered for comparison.
\end{enumerate}
The important symbols and notations used in the paper are shown in TABLE \ref{tab:title}. For simplicity, the same notations are being used to represent similar parameters of cluster processes whenever it is clear from the context. For e.g. $m$ denote the mean number of points  in a daughter point process (PP) of TCP as well as in MCP. 
\vspace{-.3em}
\subsection{Sensing Performance}
Sensing performance of a WSN can be measured in terms of Capacity functional. The capacity functional $T_\Psi(\mathsf{K})$ of Boolean process for a compact set $\mathsf{K}$ is the probability that the  the set $\K$ and $\Psi$ are not disjoint \ie
\begin{align*}
T_\Psi(\K)=\prob{\Psi\cap \K\neq \phi}.
\end{align*}
If $\set{K}$ denote any event's impact area, then $T_\Psi(\mathsf{K})$ denote the coverage/sensing probability \ie probability that the event is sensed by the sensor network. In particular, we will consider $\set{K}$ as $\Ball(\ob,r_\set{K})$  in $\mathbb{R}^2$ and denote the capacity functional for this set by $\CFB{}$. Here $r_\set{K}$ denote the size of the event. Since the network is stationary, we have taken the center at the origin $\ob$. For dynamic events, $\set{K}$ can grow in size with time \cite{pandey2018modeling}.
\begin{remark}
The capacity functional  $T_{\Psi_{\BoolP}}(\K)$ for a Boolean P process with intensity $\totallambda{\BoolP}$ and sensing range $\Ball(\ob,R)$ is given as \cite{chiu2013stochastic}	
\begin{align*}
T_{\Psi_\BoolP}(\K)&
=1-\expS{-
	\totallambda{\BoolP}
	\left|\bt(\ob,R)\oplus\set{K}\right|}.
\end{align*}
For circular set $\set{K}=\Ball(0,r_\set{K})$, $T_{\Psi_\BoolP}(\Ball(0,r_\set{K}))$ is 
\begin{align*}
\CFB{\Psi_\BoolP}&=1-\expS{-m\parentlambda\pi(R+r_\K)^2}.
\end{align*}
The probability that an arbitrary point $\set{K=\{\ob\}}$ is covered is given by:
\begin{align*}
T_{\Psi_{\BoolP}}(\{\ob\})&=1-e^{(-m\parentlambda\pi R^2)}.
\end{align*}
\end{remark}
\subsection{Power Requirement}
Power requirement $\reqEnergy{}$ of a system is defined as the sum power required by sensors in a unit area $\mathsf{u}$ to be able to communicate to their cluster head. Assuming a powerlaw path loss with path loss exponent $\alpha$ and the SNR threshold $\SThres$ required for successful communication, the power requirement is given as
\begin{align*}
\reqEnergy{}&=\expect{\sum_{\z_{ij}\in\Phi\cap \set{u}}\tau \|\z_{ij}-c_{ij}\|^{\alpha}},
\end{align*}
where $c_{ij}=\x_i$ denotes the cluster head of the sensor $\z_{ij}$.
\section{Sensing Performance}
In this section, we will derive the sensing performance of the WSN for the two considered models and provide closed form upper and lower bounds for the same.
\subsection{Boolean MC Process}
\begin{theorem}\label{thm1}
The capacity functional for the Boolean MC Process is given as (See Appendix \ref{appn A} for the proof):
\begin{align}\label{MCPCAP1}
& T_{\Psi_\BoolM}(\K)=1-\nonumber\exp\left(-\parentlambda \times \vphantom{\frac{n}{d}}\right.
\\
&\left.
\int_{\mathbb{R}^2}
\left(1-\expS{
	-\lambda_\drm\left|
	\bt(\ob,r_\drm)\cap\left(\bt(\x,R)
	\oplus \set{K}\right)
	\right|
}\right)
\dv \x\right).
\end{align}
\end{theorem}
\begin{corollary}
	For circular $\set{K}\equiv\B(\ob,r_\set{K})$, the Minkowski sum is $\B(\ob,r_\K)\oplus\B(\x,R)\equiv\B(\x,R+r_\K)$. Hence the expression for the capacity functional is:
\begin{align}\label{MCPcap}
&\CFB{\BoolM}=1-e^{\left(-2\pi\lambda_\pt \int_{0}^{r_\drm+R+r_\set{K}}
	\left(	1-\expU{-\lambda_\drm
		\areaballintersect{\ob}{x}{r_\drm}{R+r_\set{K}}
	}\right)
		x\dv x\right)}.
	\end{align}
\end{corollary}
\begin{theorem}\label{thm3}
	The upper and the lower bound for $\CFB{\BoolM}$ 
	is given as (See Appendix \ref{apndxb} for the proof.)
\begin{align*}
&
{\UCFB{\BoolM}}
=1-\exp(-\pi\lambda_\pt\Abf{r}{}{})	\exp\left(\left(
{
	\frac{\pi\lambda_\pt}{2\lambda_\drm^2\beta^2(r)}
}
\left[
-1
\right.\right.\right.\\
&\left.\left.\left.
+
2\lambda_\drm \beta(r)+e^{-
	4\lambda_\drm\beta^2(r)
}\left(
{r+r_{\drm}}{}
+|r-r_{\drm}|
e^{-
	4\lambda_\drm\beta^2(r)
}
\right)
\right]\right)\right),\\
&
{\LCFB{\BoolM}}=	1-
\exp\left(-
\pi\lambda_\pt\Abf{r}{r_\drm}{\lambda_\drm} \right)
\exp\left(
\frac{4\lambda_\pt}{\kb}\left[-2 
\vphantom{\left(\sqrt{a_{n_0}}\right)}
\right.\right.\\
&\left.\left.+2e^{-\lambda_\drm \pi \beta^2(r) }-{\pi}\sqrt{\kb}(r+r_{\drm})
\erf\left(
-\sqrt{
	\lambda_\drm \pi 
}
\beta(r) 
\right)
\right]
\right),
\end{align*}
where $r=R+r_\K$,  $\beta(r)=\min(r,r_\drm)$,	$\Abf{r}{r_\drm}{\lambda_\drm}
=(r-r_\drm)^2(1-\exp(-\lambda_\drm\pi\beta^2(r)))+4rr_\drm$.
 	
\end{theorem}
\begin{theorem} \label{thm4}
An another set of bounds for $\CFB{\BoolM}$ is
	\begin{align*}
	\UUCFB{\BoolM}&=1-\exp\left(-\pi\lambda_\pt (r_\drm+r)^2\right.
	\left.\left(1-e^{-\lambda_\drm \beta^2(r)}\right)\right),
	\\
	\LLCFB{\BoolM}&=1-\exp\left(-\pi\lambda_\pt (r_\drm-r)^2\right.\left.\left(1-e^{-\lambda_\drm \beta^2(r)}\right)\right),
	\end{align*}
	which are simpler but less tight than the ones in Theorem \ref{thm3}.
\end{theorem}

\begin{IEEEproof}	
The upper bound is derived by substituting the intersecting area $\areaballintersect{}{x}{r_\drm}{r}$
in \eqref{MCPcap} with its upper bound $\min(\pi r^2,\pi r_\drm^2)$. For lower bound, we note that from limit $x=0$ to $x=|r-r_\drm|$  $\areaballintersect{}{x}{r_\drm}{r}$ is $\min(\pi r^2,\pi r_\drm^2)$ and for $x=|r-r_\drm|$ to $x=r+r_\drm$, it can be lower bounded by $0$. Substituting these bounds in \eqref{MCPcap}, we get the lower bound.
\end{IEEEproof}
\subsubsection{Asymptotic behavior of $T_{\Psi _\BoolM}$ with $r_\drm$ while keeping $\totallambda{\BoolM}$ fixed.}
By increasing the  $r_\drm$ while  keeping $\totallambda{\BoolM}$ fixed, we can decrease the clustering of points and spread points more in the space. Hence, the asymptotic behavior of $T_{\Psi _\BoolM}$ helps us in understanding the impact of clustering (or mutual-attraction of points) on the sensing performance.

\textbf{When $r_\drm\rightarrow 0$:}  Taking the limit in the Theorem \ref{thm4},
we get the lower bound $\LLCFB{\BoolM}$
\begin{align*}
&=1-\lim_{r_\drm\rightarrow 0}\exp\left(-\pi\lambda_\pt (r_\drm-r)^2\left(1-e^{-\lambda_\drm \min(\pi r_\drm^2,\pi r^2)}\right)\right),\\
&=1-\exp\left(-\pi\lambda_\pt r^2\left(1-e^{-m}\right)\right).
\end{align*}
Similarly, taking limit of the upper bound in Theorem \ref{thm3}, we get  
$\UCFB{\BoolM}$
\begin{align*}
&=1-\lim_{r_\drm\rightarrow 0}\exp\left(-\pi\lambda_\pt (r_\drm+r)^2\left(1-e^{-\lambda_\drm \min(\pi r_\drm^2,\pi r^2)}\right)\right),\\
&=1-\exp\left(-\pi\lambda_\pt r^2\left(1-e^{-m}\right)\right).
\end{align*}
Since, both upper and lower bounds converge to the same function, the capacity functional converges to the same function. As $r_\drm \rightarrow 0$, $\CFB{\BoolM}$ converges to $\CFB{\BoolP}$ with intensity $\lambda_{\pt}(1-e^{-m})$.

\textbf{When $r_\drm \rightarrow\infty$:} 
{Using the expressions from Theorem \ref{thm3}, we see that both the upper and lower bound converge to $1-\exp(-m\pi\lambda_{\pt}r^2)$. Note that this is equal to the capacity functional of a Boolean P process with intensity $m\lambda_{\pt}$.}
The required power for Boolean MC process is given as: 
\begin{align*}
\reqEnergy{\BoolM}&=\expect{\sum_{\z_{ij}\in\Phi\cap \set{u}}\tau \|\z_{ij}-c_{ij}\|^{\alpha}},
\\
&=\totallambda{\BoolM}\int_{\set{u}} \tau \palmexpectx{\|\z-c_{\z}\|^{\alpha}}{\z}\dd\z
=\totallambda{\BoolM} \tau \palmexpect{\|y\|^{\alpha}},\\
&=m\parentlambda \tau \frac1{\alpha/2+1} r_\drm^\alpha.
\end{align*}
\subsection{Boolean TC Process}
\begin{theorem}\label{thm5}
	The capacity functional for a Boolean TC process is given by $\CFB{\BoolT}$ (For proof see Appendix \ref{prf:thm5} )
	\begin{align*}\label{TCPcapacity1}
&=1-\exp\left(-2\pi\lambda_\pt\int_{x=0}^{\infty}\left(1-\vphantom{\frac24}\right.\right.\nonumber\\
&\left.\left.\exp\left(-\frac{m}{2\pi\sigma^2}\int_{\theta=0}^{2\pi}\int_{t=0}^{r}e^{-\frac{x^2+t^2+2xt\cos\theta}{2\sigma^2}}t\dv t\dv \theta\right)\right)x\dv x\right),
	\end{align*}
	where $r=R+r_\K$.
\end{theorem}

The required power for Boolean TC process is given as
\begin{align*}
\reqEnergy{\BoolT}&=\expect{\sum_{\z_{ij}\in\Phi\cap \set{u}}\tau \|\z_{ij}-c_{ij}\|^{\alpha}}=\totallambda{\BoolT}\int_{\set{u}} \tau \palmexpectx{\|\z-c_{\z}\|^{\alpha}}{\z}\dd\z,
\\
&=\totallambda{\BoolT} \tau \palmexpect{\|y\|^{\alpha}}=m\parentlambda \tau \Gamma(\alpha/2+1) (2\sigma^2)^{\alpha/2}.
\end{align*}
\subsection{Boolean Poisson Process}
In this case, the required power is given as
\begin{align*}
\reqEnergy{}&=\expect{\sum_{\z_{ij}\in\Phi\cap \set{u}}\tau \|\z_{ij}-c_{ij}\|^{\alpha}},\\
&=\totallambda{\BoolP}\int_{\set{u}} \tau \palmexpect{\|c_{0}\|^{\alpha}}\dd\z
=\totallambda{\BoolP} \tau \palmexpect{\|c_{0}\|^{\alpha}}.
\end{align*}
Now, $c_0$ is the closest cluster head (which is modeled as PPP with intensity $\parentlambda$), hence the distribution of its distance from the origin is given as
\[f_{\|c_0\|}(c)=2\pi\parentlambda{} c\expS{-\pi\parentlambda{} c^2}.
\]
Hence, 
\begin{align*}
\reqEnergy{}&=
\totallambda{\BoolP} 2\pi\parentlambda\tau 
\int c^{\alpha} c\expS{-\pi\parentlambda c^2}\dd c,
\\
&=m\parentlambda {(\pi\parentlambda)}^{-\alpha/2}\SThres \Gamma(\alpha/2+1).
\end{align*}
\section{Coverage Analysis with Sensor Power Constraints}
Let the required per-unit area power to be the $\mathcal{E}_{\mathrm{net}}$ which is kept constant for all three deployments. Now, for the provided $\mathcal{E}_{\mathrm{net}}$, we will derive the parameter specifications for the three models to be able to compare their coverage performance.
\subsection{Boolean MC process}
In case of Boolean MC process, the cluster radius should be equal to
\begin{align*}
r_\drm&=\left[{(\mathcal{E}_{\mathrm{net}}(1+.5\alpha))}/{m\lambda_\pt\tau}\right]^{\frac{1}{\alpha}}.
\end{align*}
\subsection{Boolean TC process}
 The variance parameter $\sigma$ to achieve required per-unit area power $\mathcal{E}_{\mathrm{net}}$ should be equal to
\begin{align*}
\sigma&=\left[{(\mathcal{E}_{\mathrm{net}})}/{m\lambda_\pt\tau\Gamma(1+.5\alpha)2^{.5\alpha}}\right]^{\frac{1}{\alpha}}.
\end{align*}
\subsection{Boolean P process}
The mean number of points per cluster head in Boolean P process should be equal to
\begin{align*}
m={\mathcal{E}_{\mathrm{net}}}/{(\Gamma(.5\alpha+1)\tau\lambda_{\pt}(\pi\lambda_{\pt})^{-.5\alpha})}.
\end{align*}
By fixing $m$, we can get the expression for the coverage probability in Boolean P process.
\begin{figure}[ht!]
\def\svgwidth{\textwidth}
	\centering
	\includegraphics[width=.46\textwidth]{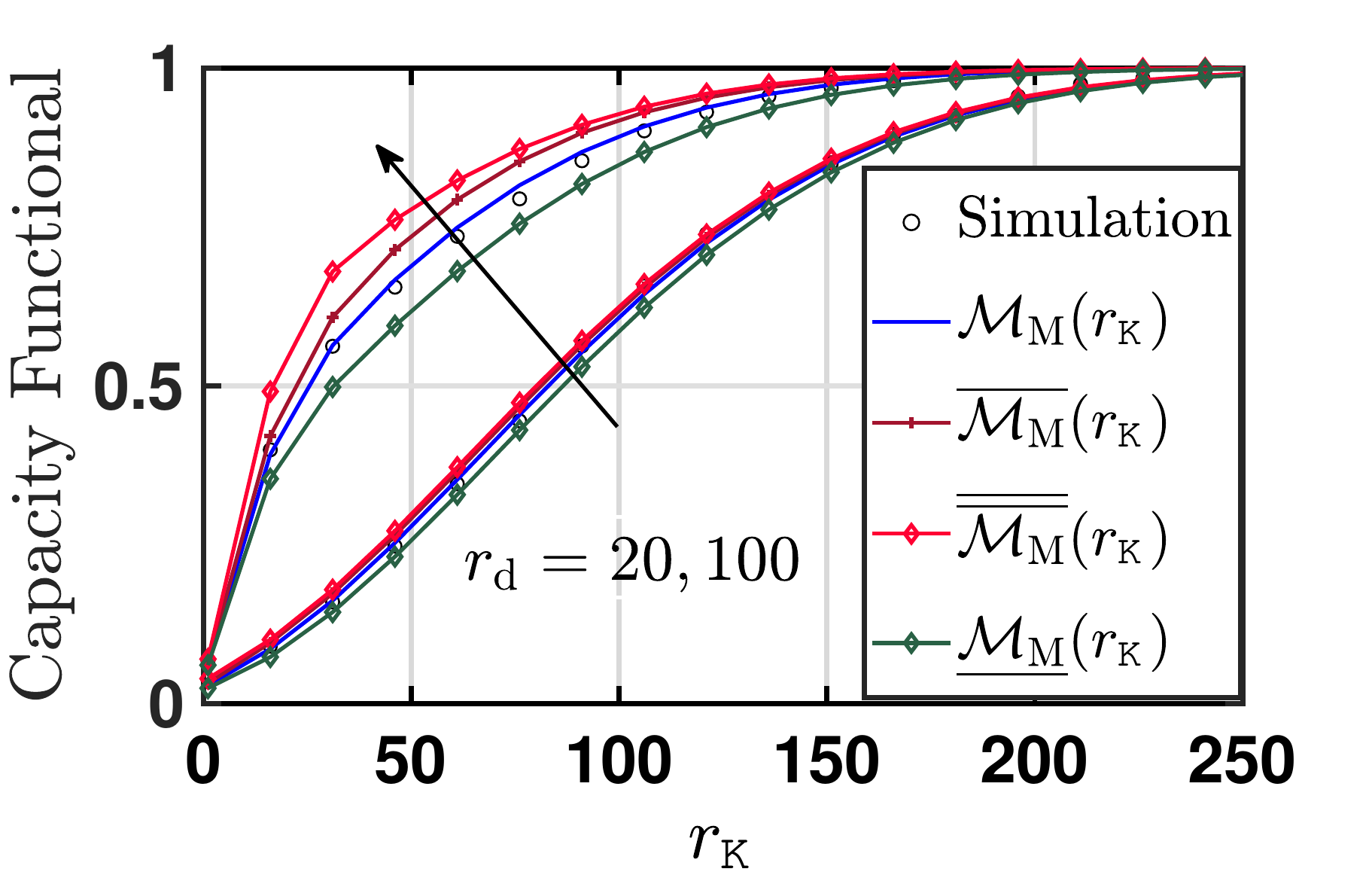}
	\caption{Capacity functional/sensing probability  vs event size $r_\set{K}$ for Boolean MC Process. Here $m=30$. 
	To increase total covered area, sensors should be less clustered (larger $r_\drm$).} 
	 \label{capcityfunctionalwithfixed}
	\def\svgwidth{\textwidth}
	\centering
	\includegraphics[width=.46\textwidth]{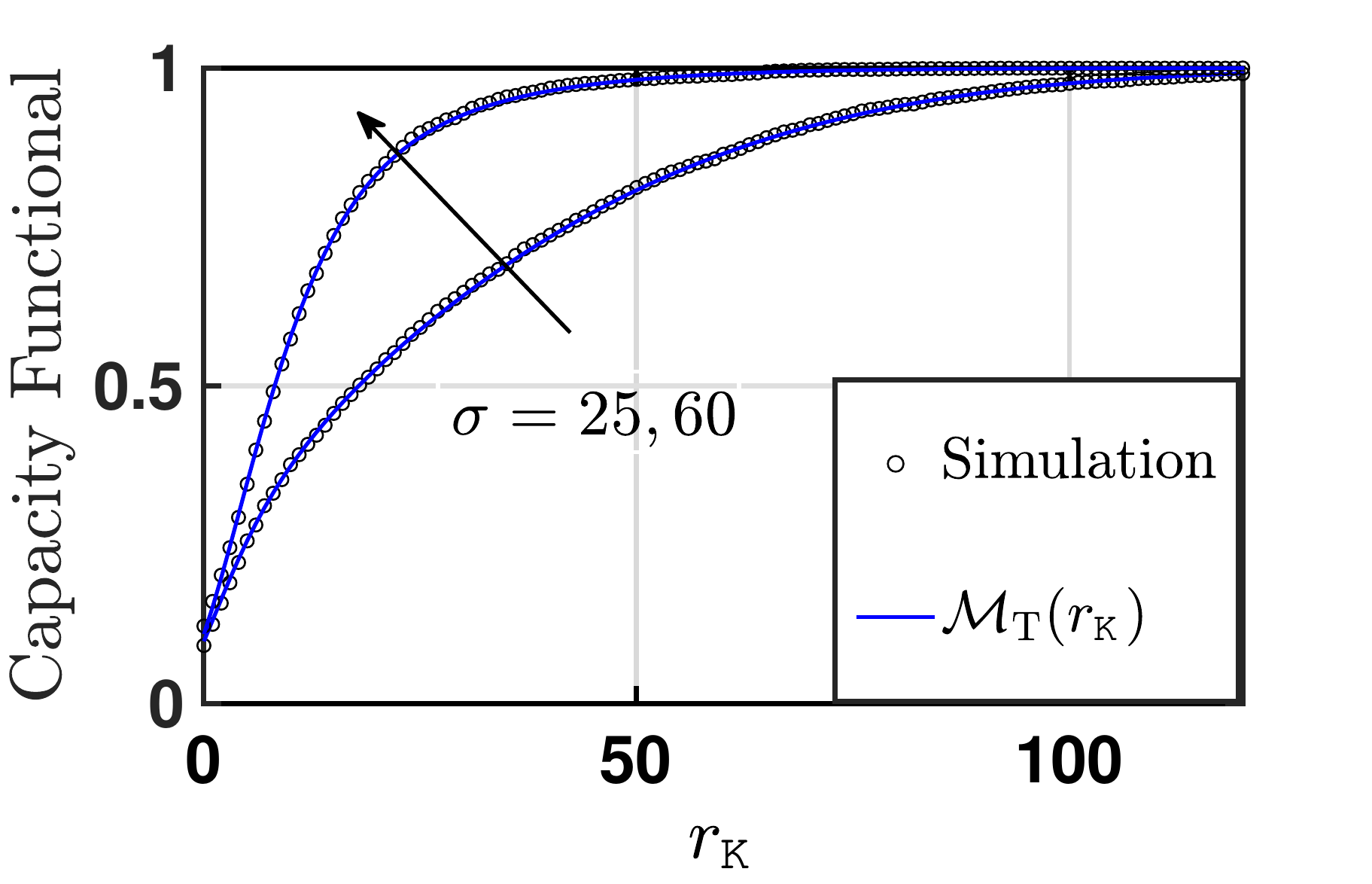}
	\caption{ Capacity functional for Boolean TC process with $R=5$ and $m=30$. Increasing variance increases the capacity functional as the daughter points are distantly located with each other in each cluster and have higher chance to cover distinct areas. } \label{TCPcapacity}
		\vspace{-.5em}
\end{figure}
\begin{figure}[ht!]
	\def\svgwidth{\textwidth}
	\centering
	\includegraphics[width=.46\textwidth]{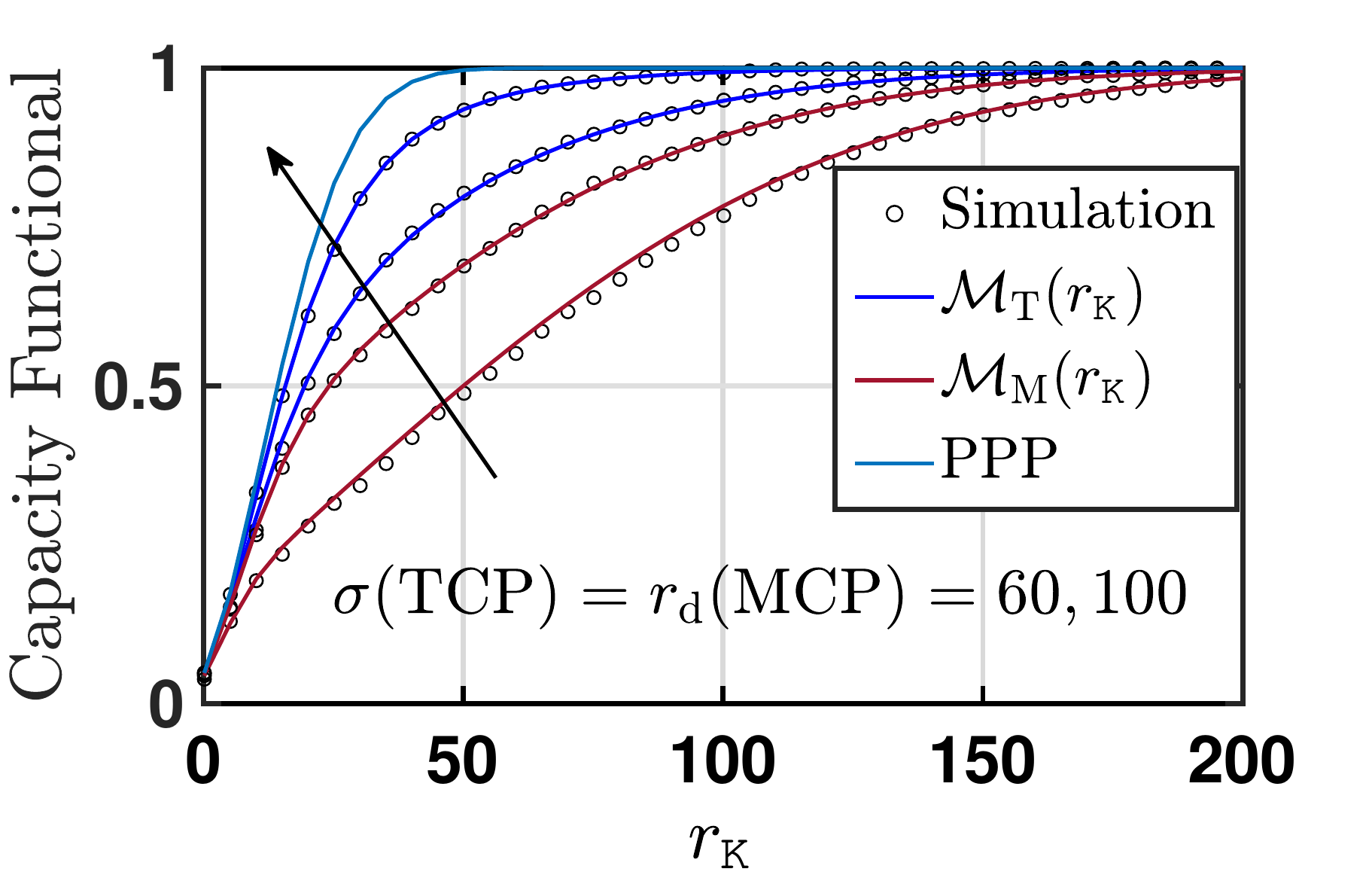}
	\caption{ Capacity functional for all three processes with increasing $r_\drm(\Psi_\BoolM)$ and $\sigma(\Psi_\BoolT)$. The mean number of points $m=30$ in the daughter point process  and the sensing range $R=5$ is kept fixed. For $\Psi_\BoolP$, the intensity is  $\totallambda{\BoolP}=m\lambda_\pt$.} \label{comparision}
	\def\svgwidth{\textwidth}
	\centering
	\includegraphics[width=.46\textwidth]{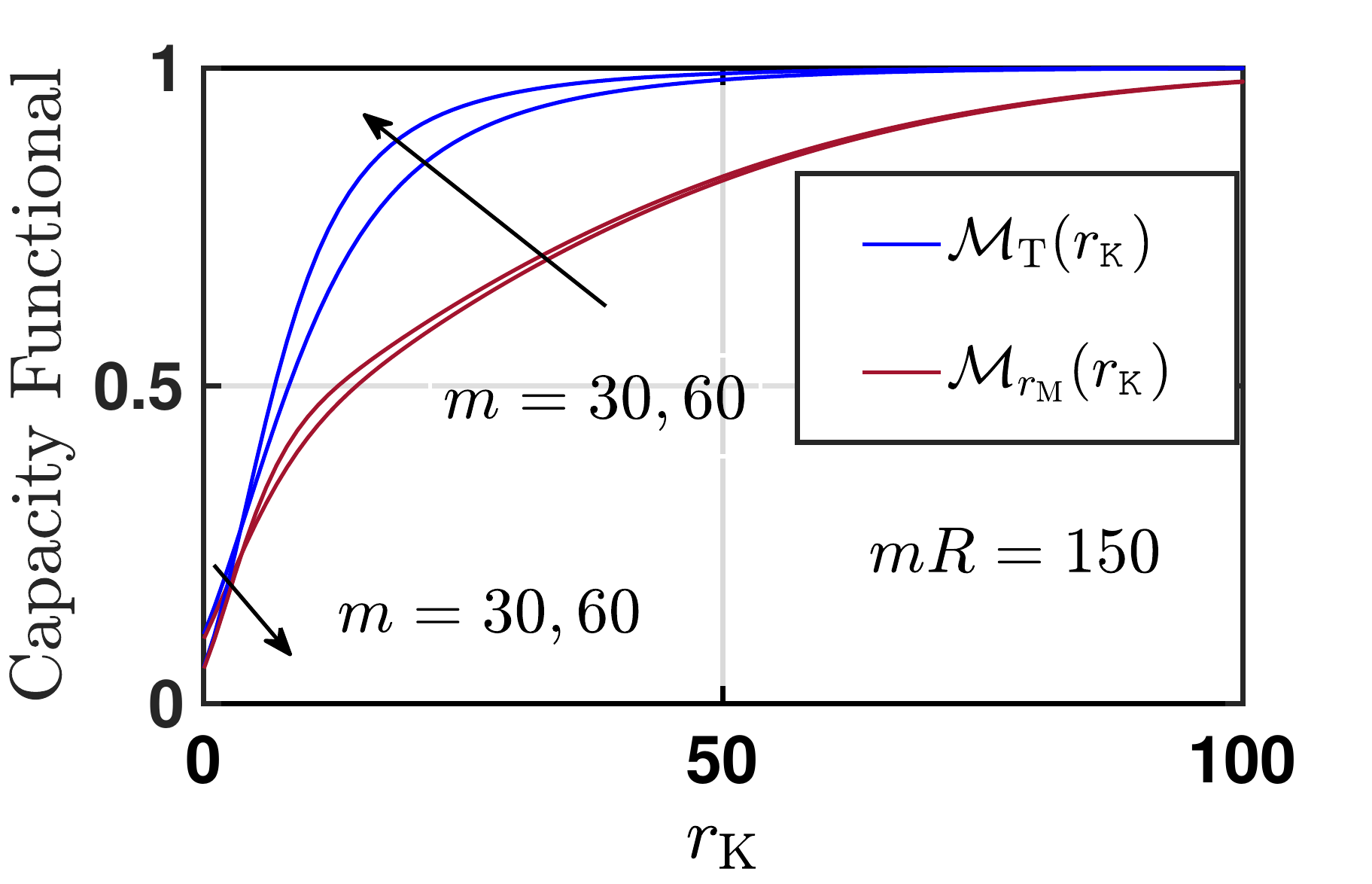}
	\caption{Variation of capacity functional with $m$, while keeping the $m\times R=150$ fixed. Here $\parentlambda=50/km^2$. It is better to deploy more sensor with smaller sensing range.} \label{capcityfunvary}
	\def\svgwidth{\textwidth}
	\centering
	\includegraphics[width=.46\textwidth]{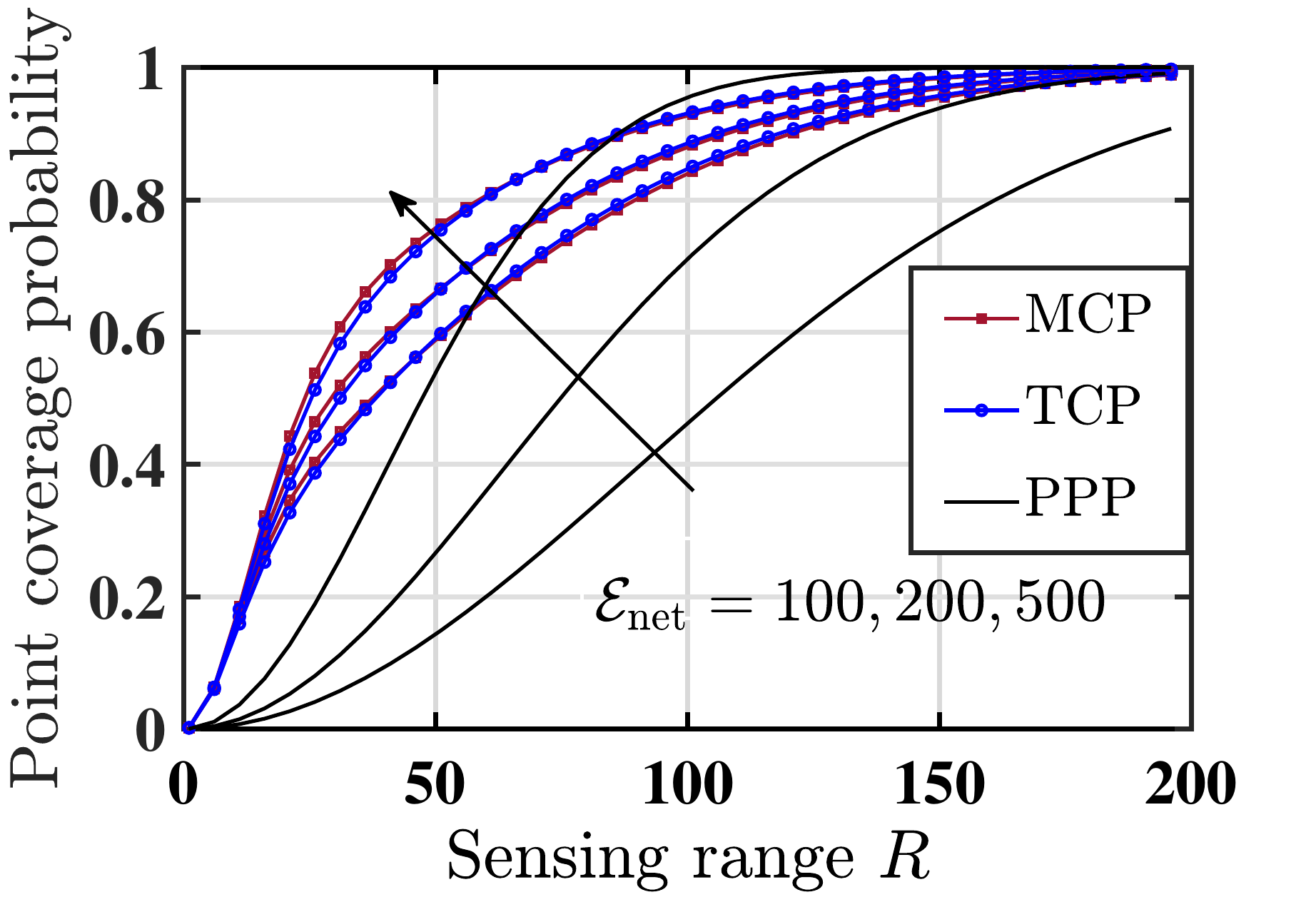}
	\caption{Variation of  point coverage probability. { Boolean-cluster model have better point coverage probability when sensors have limited power. }} \label{pointcoverage}
\end{figure}
\\ \textbf{Capacity functional/sensing probability of Boolean-TC process:}
Fig. \ref{TCPcapacity} shows the variation of sensing probability $\CFB{\BoolT}$ with respect to the event size $r_\K$ for two values of cluster spread $\sigma$. Similar trends as Boolean MC process can be seen here.\\
\textbf{Impact of sensors' deployment:} 
Fig. \ref{comparision} shows the comparison  among the sensing performance of the three deployments $\Psi_{\BoolM}$,  $\Psi_{\BoolT}$ and  $\Psi_{\BoolP}$. For the fair comparison, we took $\sigma$ of  $\Psi_{\BoolT}$ equal to $r_\drm$ of  $\Psi_{\BoolM}$. Recall that  $\Psi_{\BoolM}$ has highest clustering,  $\Psi_{\BoolT}$ has moderate and  $\Psi_{\BoolP}$ has no clustering (independence across sensors). It can be observed that  $\Psi_{\BoolM}$ has the lowest capacity functional. It can be justified by the fact that the confinement of daughter points inside a ball in  $\Psi_{\BoolM}$ increases the overlap among sensing regions of sensors in the cluster. 
In case of  $\Psi_{\BoolT}$, the daughter points are more scattered and can cover a larger region. In the case of  $\Psi_{\BoolP}$, points are the most scattered which leads to the highest performance. The graph also shows the variation of capacity functional over $\sigma$ and  $r_\drm$. The graph depicts that increasing these parameter,  $\Psi_{\BoolM}$ or  $\Psi_{\BoolT}$ will converge to   $\Psi_{\BoolP}$ (and hence their sensing performance). 
\\ 
\textbf{Trade-off between number of sensors vs sensing radius:}
Fig. \ref{capcityfunvary} shows the variation of capacity functional with $m$, while keeping  $m\times R=150$ fixed.  $m\times R$ serves as a proxy to the system cost as increasing any of $m$ and $R$ increases the cost (both- infra stricture and operating). Our analysis shows that for  small event size $r_\K$, a WSN with larger $R$ provides higher sensing performance. However, as $r_\set{K}$ increases, it is better to have a higher number of sensors than the larger sensing range.\\

\textbf{Impact on coverage with constrained network power:}
The point coverage probability is defined as the capacity functional with $r_\K=0$. For the cluster processes we have considered $m=30$ and $\lambda_{\pt}=20\times10^{-6}$. We now fix the per-cluster power requirement $\mathcal{E}_{\mathrm{net}}$ and compute coverage probability of the three deployments. Fig. \ref{pointcoverage} shows the coverage probability of three networks under various power levels. At lower $\mathcal{E}_{\mathrm{net}}$, it can be observed that MCP and TCP Boolean models provide better coverage probability. {As power constraints are less restrictive, the coverage probability increases. With sensors having higher power, the PPP Boolean can provide better coverage. We can observe that clustered deployments can provide better coverage under stricter power  constraints.}
\section{Conclusion}
This paper performs the coverage analysis of a wireless sensor network when the sensors' location are according to PPP, MCP and TCP. In these cases, sensor network can be modeled via Boolean-P, Boolean MCP and Boolean TCP processes.  We derived the expressions for the
capacity function of the two clustered deployments. As far as highest coverage is the goal, PPP performs the best of the three as the rest of two processes are attractive processes and this difference in the coverage
area of clustered and PPP deployments reduces with the sensing radius of  individual sensors. However, clustered deployments require less energy as their clustered heads are statistically closer than that of Boolean-P deployment. We also derive the average per-cluster required power to achieve a certain coverage  area. Raising the average per cluster power allows larger the cluster radius and thus higher point coverage probability. A general trade-off can be see between per-cluster required energy and coverage area when choosing between clustered or Poisson deployments. It is also observed that when sensors have low power levels, deployment, according to a Boolean cluster process, can provide better performance compared to deployment according to a Boolean
Poisson process.
\appendices
\section{} \label{appn A}
The probability that $\K$ does not intersect with the covered region $\Psi_\BoolM$ is given by: $\mathbb{P}(\Psi_\BoolM\cap \K\neq \phi)=$
\begin{small}
	\begin{align*}
	&\mathbb{E}\left[\prod_{\x_i\in\Phi_{\pt}}\prod_{\y_j^{(i)}\in \Phi_{\drm}^{(i)}}\left(
	\1\left(
	(\x_i+\y_j^{(i)}+S_j^{(i)})
	\cap
	\K 
	\right)\neq \phi
	\right)\right],\\
	&=1-\mathbb{E}\left[\prod_{\x_i\in\Phi_{\pt}}\prod_{\y_j^{(i)}\in \Phi_{\drm}^{(i)}}
	\1\left(
		(\x_i+\y_j^{(i)}+S_j^{(i)})\cap \K = \phi
	\right)
	\right],\\
			&\stackrel{(a)}=1-\mathbb{E}
\left[\prod_{\x_i\in\Phi_{\pt}}
\prod_{\y_j^{(i)}\in \Phi_{\drm}^{(i)}}
\1\left(\y_j^{(i)}\notin \left((-\x_i+S_{j}^{(i)})\oplus \K
\right) \right)\right],\\
		&\stackrel{(b)}=1-\mathbb{E}_{\Phi_{\pt}}\left[\prod_{\x_i\in \Phi_{\pt}}\exp(-\lambda_\drm|\bt(\ob,r_\drm)\cap\bt(-\x_i,R)
	\oplus \K |) \right].
	\end{align*}
\end{small}
Here  $(a)$ is due to the definition of Minkowski sum and {$(b)$ is applying the PGFL of $\Phi_\drm^{(i)}$. Applying the PGFL of $\Phi_{\pt}$  we get the MCP capacity functional \eqref{MCPCAP1}.}
\section{} \label{apndxb}
\begin{figure}[ht!]
	\def\svgwidth{\textwidth}
	\centering
	\includegraphics[width=.46\textwidth]{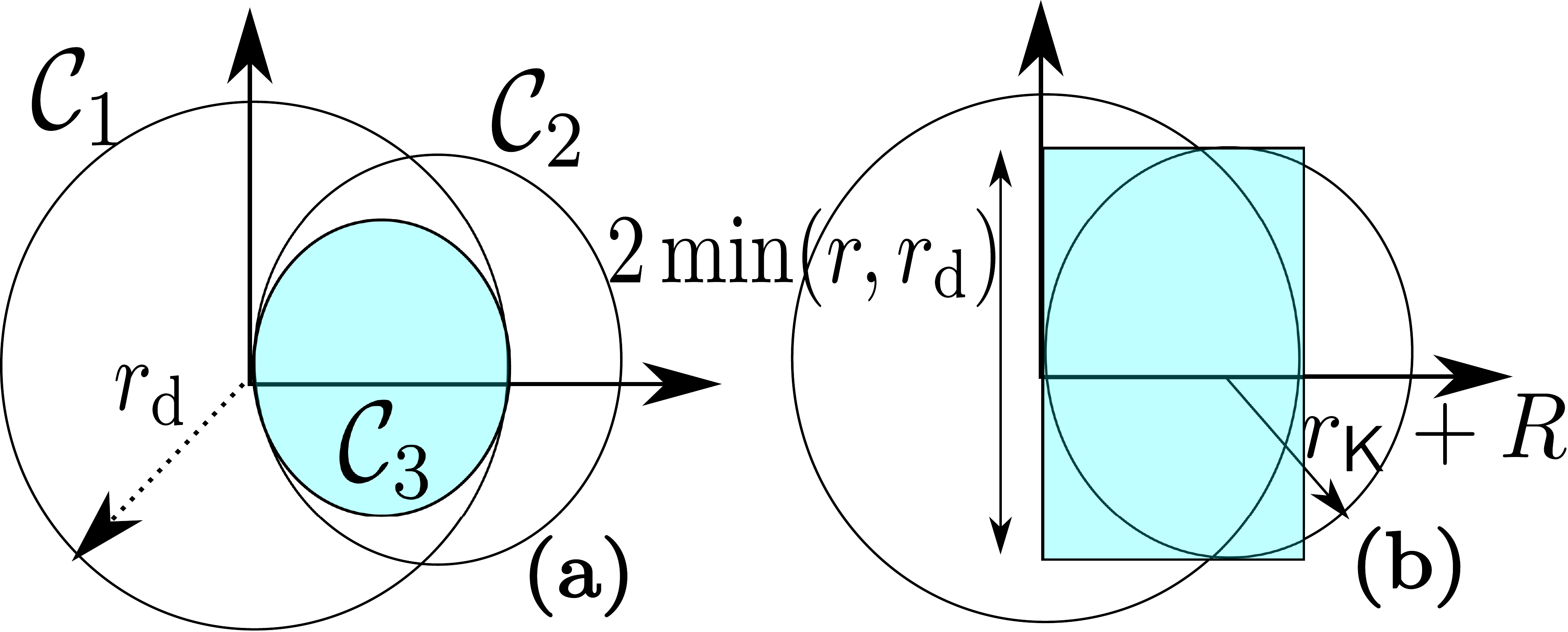}
	\caption{ Illustration showing bounds on the intersecting region between the two circle. Geometrical shapes can be used to lower bound  (circle)  and upper bound (rectangle) the intersecting region.} \label{circleintersection}
		\vspace{-.5em}
\end{figure}

To solve the integral expression presented in \eqref{MCPcap} we are presenting couple of bounds over the intersecting region $\areaballintersect{}{x}{r_\drm}{r}$. The bounding techniques are similar to \cite[Th. 2, App. C]{pandey2019detection}. Fig. \ref{circleintersection} depicts the shapes (circle and rectangle) which can bound the area of intersection between the two circle.  Let the two circle be $\C_1\equiv \B(\ob,r_\drm)$  and $\C_2\equiv\B(\ob,r_{\mathsf{K}}+R)$, of radius $r_{\drm}$ and $r_{\mathsf{K}}+R$ respectively. A third circle $\C_3$ of radius $\frac{r+r_{\drm}-x}{2}$ centered at $(\frac{r-r_{\drm}+x}{2},0)$ will be entirely inside the intersecting region. Similarly, a rectangle of width $r+r_\drm-x$ and height $2\min(r,r_\drm)$ will completely cover the intersecting area hence acts as an upper bound. For detailed discussion over the bounds readers are advised to refer \cite{pandey2019detection,pandey2019contact}.
\section{}\label{prf:thm5}
The proof is similar to the proof in Appendix \ref{appn A}. For Boolean TCP process, the null probability $\mathbb{P}(\Psi_{\BoolT}\cap \K\neq \phi)$ is 
\begin{small}
	\begin{align*}
	&=\mathbb{E}\left[\prod_{\x_i\in\Phi_{\pt}}\prod_{\y_j^{(i)}\in \Phi_{\drm}^{(i)}}\left(\1\left((\x_i+\y_j^{(i)}+S_j^{(i)})\cap \K \right)\neq \phi\right)\right],\\
&=1-\mathbb{E}_{\Phi_{\pt}}\left[\prod_{\x_i\in \Phi_{\pt}}\expectop_{\y_j^{(i)}|\x_i}\left[\prod_{\y_j^{(i)}\in \mathbb{R}^2}\1(\y_j^{(i)}\notin\bt(-\x_i,R)\oplus \K )\right]\right],\\
	&\stackrel{(a)}=1-\mathbb{E}_{\Phi_{\pt}}\left[\prod_{\x_i\in \Phi_{\pt}}\int_{\y\in\bt(\x_i,R+r_\K)}\frac{m}{2\pi\sigma^2}\exp\left(-\frac{y^2}{2\sigma^2}\right)y\dv y\dv \theta\right].
	\end{align*}
\end{small}
{Here $(a)$ is acquired by the PGFL of $\Phi_{\drm}^{(i)}$}. Let $r=R+r_\K$ and $\y=\x_i+\tf \implies \dv\y=\dv \tf$:
\begin{small}
	\begin{align}\label{last_step}
	&=1-\mathbb{E}_{\Phi_{\pt}}\left[\prod_{\x_i\in \Phi_{\pt}}\int_{\tf\in\bt(\ob,r)}\frac{m}{2\pi\sigma^2}e^{\left(-\frac{||\x_i+\tf||^2}{2\sigma^2}
		\right)}\dv \tf\right].
	\end{align}
\end{small}
Without loss of generality, it can be assumed that $\x_i\equiv(x_i,0)$ as the parent point process $\Phi_{\pt}$ is rotation invariant. Thus, replacing $||\x+\tf||^2$ with $x^2+t^2+2xt\cos\theta$. {
	Applying the PGFL of $\Phi_{\pt}$ in \eqref{last_step}, we get the Theorem \ref{thm5}.} 

%
%

\ifCLASSOPTIONcaptionsoff
  \newpage
\fi

\bibliographystyle{ieeetran}



\begin{thebibliography}{10}
\providecommand{\url}[1]{#1}
\csname url@samestyle\endcsname
\providecommand{\newblock}{\relax}
\providecommand{\bibinfo}[2]{#2}
\providecommand{\BIBentrySTDinterwordspacing}{\spaceskip=0pt\relax}
\providecommand{\BIBentryALTinterwordstretchfactor}{4}
\providecommand{\BIBentryALTinterwordspacing}{\spaceskip=\fontdimen2\font plus
\BIBentryALTinterwordstretchfactor\fontdimen3\font minus
  \fontdimen4\font\relax}
\providecommand{\BIBforeignlanguage}[2]{{%
\expandafter\ifx\csname l@#1\endcsname\relax
\typeout{** WARNING: IEEEtran.bst: No hyphenation pattern has been}%
\typeout{** loaded for the language `#1'. Using the pattern for}%
\typeout{** the default language instead.}%
\else
\language=\csname l@#1\endcsname
\fi
#2}}
\providecommand{\BIBdecl}{\relax}
\BIBdecl

\bibitem{iyengar2016distributed}
S.~S. Iyengar and R.~R. Brooks, \emph{Distributed Sensor Networks: Sensor
  Networking and Applications (Volume Two)}.\hskip 1em plus 0.5em minus
  0.4em\relax CRC press, 2016.

\bibitem{haenggibook}
M.~Haenggi, \emph{Stochastic geometry for wireless networks}.\hskip 1em plus
  0.5em minus 0.4em\relax Cambridge University Press, 2012.

\bibitem{liu2004study}
{Benyuan Liu} and D.~{Towsley}, ``A study of the coverage of large-scale sensor
  networks,'' in \emph{in Proc. International Conference on Mobile Ad-hoc and
  Sensor Systems}, Oct 2004, pp. 475--483.

\bibitem{baek2007spatial}
S.~J. {Baek} and G.~{de Veciana}, ``Spatial model for energy burden balancing
  and data fusion in sensor networks detecting bursty events,'' \emph{IEEE
  Trans. Inf. Theory}, vol.~53, no.~10, pp. 3615--3628, Oct 2007.

\bibitem{BaccelliBook}
F.~Baccelli and B.~B{\l}aszczyszyn, \emph{Stochastic Geometry and Wireless
  Networks}, 2nd~ed.\hskip 1em plus 0.5em minus 0.4em\relax NOW Publishers,
  2009, vol.~1.

\bibitem{chiu2013stochastic}
S.~N. Chiu, D.~Stoyan, W.~S. Kendall, and J.~Mecke, \emph{Stochastic geometry
  and its applications}.\hskip 1em plus 0.5em minus 0.4em\relax John Wiley \&
  Sons, 2013.

\bibitem{flint2017wireless}
I.~{Flint}, H.~{Kong}, N.~{Privault}, P.~{Wang}, and D.~{Niyato}, ``Wireless
  energy harvesting sensor networks: Boolean–poisson modeling and analysis,''
  \emph{IEEE Trans. Wireless Commun.}, vol.~16, no.~11, pp. 7108--7122, Nov
  2017.

\bibitem{pandey2018modeling}
\BIBentryALTinterwordspacing
K.~Pandey and A.~Gupta, ``Modeling and analysis of wildfire detection using
  wireless sensor network with {P}oisson deployment,'' \emph{in Proc. IEEE
  ANTS, Dec. 2018}. [Online]. Available: \url{https://arxiv.org/abs/1810.07511}
\BIBentrySTDinterwordspacing

\bibitem{mekikis2018connectivity}
P.~{Mekikis}, E.~{Kartsakli}, A.~{Antonopoulos}, L.~{Alonso}, and
  C.~{Verikoukis}, ``Connectivity analysis in clustered wireless sensor
  networks powered by solar energy,'' \emph{IEEE Trans. Wireless Commun.},
  vol.~17, no.~4, pp. 2389--2401, April 2018.

\bibitem{pandey2019contact}
K.~{Pandey}, H.~S. {Dhillon}, and A.~K. {Gupta}, ``On the contact and
  nearest-neighbor distance distributions for the n-dimensional cluster
  process,'' \emph{IEEE Wireless Commun. Lett}, pp. 1--1, 2019.

\bibitem{afshang2017nearesttcp}
M.~Afshang, C.~Saha, and H.~S. Dhillon, ``Nearest-neighbor and contact distance
  distributions for {T}homas cluster process,'' \emph{IEEE Commun. Lett},
  vol.~6, no.~1, pp. 130--133, 2017.

\bibitem{last1999empty}
G.~Last and M.~Holtmann, ``On the empty space function of some germ-grain
  models,'' \emph{Pattern Recognition}, vol.~32, no.~9, pp. 1587--1600, 1999.

\bibitem{pandey2019detection}
\BIBentryALTinterwordspacing
K.~K. Pandey and A.~K. Gupta, ``On detection of critical events in a finite
  forest using randomly deployed wireless sensors,'' \emph{in Proc. SpasWin
  June 2019}. [Online]. Available: \url{https://arxiv.org/pdf/1904.09543.pdf}
\BIBentrySTDinterwordspacing

\end{thebibliography}
\vspace{12pt}




\end{document}